\documentclass{article}%
\usepackage{amsmath}
\usepackage{amsfonts}
\usepackage{amssymb}
\usepackage{graphicx}%
\providecommand{\U}[1]{\protect\rule{.1in}{.1in}}
\newtheorem{theorem}{Theorem}

\newtheorem{proposition}[theorem]{Proposition}

\begin{document}

\title{On the Motion of a Free Particle in the de Sitter Manifold}
\author{Waldyr A. Rodrigues Jr. and Samuel A. Wainer\\Institute of Mathematics Statistics and Scientific Computation\\IMECC-UNICAMP\\e-mail: walrod@ime.unicamp.br~~~~samuelwainer@ime.unicamp.br}
\date{January 21 2016}
\maketitle

\begin{abstract}
Let $M=SO(1,4)/SO(1,3)\simeq S^{3}\times\mathbb{R}$ (a parallelizable
manifold) be a submanifold in the structure $(\mathring{M}%
,\boldsymbol{\mathring{g}})$ (hereafter called the bulk) where $\mathring
{M}\simeq\mathbb{R}^{5}$ and $\boldsymbol{\mathring{g}}$ is a pseudo Euclidian
metric of signature $(1,4)$. Let $\boldsymbol{i}:M\rightarrow\mathbb{R}^{5}$
be the inclusion map and let \ $\boldsymbol{g}=\boldsymbol{i}^{\ast
}\boldsymbol{\mathring{g}}$ be the pullback metric on $M$. It has signature
$(1,3)$ Let $\boldsymbol{D}$ be the Levi-Civita connection of $\boldsymbol{g}%
$. We call the structure $(M,\boldsymbol{g})$ a de Sitter manifold and
$M^{dSL}=(M=\mathbb{R\times}S^{3},\boldsymbol{g},\boldsymbol{D},\tau
_{\boldsymbol{g}},\uparrow)$ a de Sitter spacetime structure, which is \ of
course orientable by $\tau_{\boldsymbol{g}}\in\sec%
{\textstyle\bigwedge\nolimits^{4}}
T^{\ast}M$ and time orientable (by $\uparrow$).\ Under these conditions we
prove that if the motion of a free particle moving on $M$ happens with
constant \emph{bulk} angular momentum then its motion in the structure
$M^{dSL}$ is a timelike geodesic. Also any geodesic motion in the structure
$M^{dSL}$ implies that the particle has constant angular momentum in the bulk.

\end{abstract}

\section{Introduction}

In what follows $SO(1,4)$ and $SO(1,3)$ denote the special pseudo-orthogonal
groups in $\mathbb{R}^{1,4}=\mathbb{(}\mathring{M}=\mathbb{R}^{5}%
,\boldsymbol{\mathring{g}})$ where $\boldsymbol{\mathring{g}}$ is a metric of
signature $(1,4)$. The \emph{de Sitter manifold}.$M$ can be viewed as a brane
(a submanifold) in the structure $\mathbb{R}^{1,4}$. The structure
$M^{dSL}=(M,\boldsymbol{g},\boldsymbol{D},\tau_{\boldsymbol{g}},\uparrow)$
will be called\emph{ Lorentzian de Sitter spacetime structure} where, if
$\boldsymbol{\iota}:\mathbb{R\times}S^{3}\rightarrow\mathbb{R}^{5}$ is the
inclusion mapping, $\boldsymbol{g=\iota}^{\ast}\boldsymbol{\mathring{g}}$ and
$\boldsymbol{D}$ is the parallel projection on $M$ of the pseudo Euclidian
metric compatible connection $\boldsymbol{\mathring{D}}$ in $\mathbb{R}^{1,4}$
(details in \cite{rw2013,rw2015}). As well known, $(M,\boldsymbol{g)}$, a
pseudo-sphere is a spacetime of constant Riemannian curvature. It has ten
Killing vector fields. The Killing vector fields are the generators of
infinitesimal actions of the group $SO(1,4)$ (called the de Sitter group) in
$M$. The group $SO(1,4)$ acts transitively\footnote{A group $G$ of
transformations in a manifold $M$ ($\sigma:G\times M\rightarrow M$ by
$(g,x)\mapsto\sigma(g,x)$) is said to act transitively on $M$ if for
arbitraries $x,y\in M$ there exists $g\in G$ such that $\sigma(g,x)=y$.} in
$SO(1,4)/SO(1,3)$, which is thus a homogeneous space (for $SO(1,4)$).

The structure $M^{dSL}$ has been used by many physicists as an alternative
arena for the motion of particles and fields in place of the Minkowski
spacetime structure\footnote{Minkowski spacetime is the structure
$\mathfrak{M=(\mathcal{M}}=\mathbb{R}^{4},\boldsymbol{\eta},D,\tau
_{\boldsymbol{\eta}},\uparrow\mathfrak{)}$ where $\boldsymbol{\eta}$ is the
usual Minkowski metric, $\tau_{\boldsymbol{\eta}}\in\sec%
{\textstyle\bigwedge\nolimits^{4}}
T^{\ast}\mathcal{M}$ defines an orientation and $\uparrow$ denotes that
\ $(\mathfrak{\mathcal{M}},\boldsymbol{\eta})$ is time orientable. Details in
\cite{rodcap2007}.} $\mathfrak{M}$. One of the reasons is that the isometry
group of the structure $(M,\boldsymbol{g})$ is the de Sitter group, which as
well known reduces to the Poincar\'{e} group when he radius $\ell$ of
($M,\boldsymbol{g})$ goes to $\infty$. Now, as well known the natural motion
of a free particle of mass $m$ in $\mathfrak{M}$ occurs with
constant\ momentum $\boldsymbol{p}=m\varkappa_{\ast}$ where $\varkappa
:\mathbb{R\rightarrow}\mathcal{M}$ is a timelike curve pointing to the future.
The question which naturally arises is the following:

\begin{center}
\emph{Which is the natural motion of a free particle of mass }$m$\emph{ in the
structure }$(M,g)$\emph{? }
\end{center}

One natural suggestion given the well known relation between the de Sitter and
Poincar\'{e} groups \cite{gursey} is that such a motion occurs with constant
angular momentum $\boldsymbol{L}$ as determined by (hyper observers) living in
the bulk. Given this hypothesis we prove the following proposition:
(a)\emph{:} If a particle travels with geodesic motion in the structure
$M^{dSL}$ then its bulk angular momentum $\boldsymbol{L}$ is constant.
(b)\emph{:} Also, if the motion of a particle of mass $m$ constrained to live
in $M$ occurs with constant bulk angular $\boldsymbol{L}$ then its motion for
an observer living in the brane $M$ is described by a timelike geodesic in the
structure $M^{dSL}$.

The paper is composed of three sections. Section 2, called preliminaries fixes
the necessary notations and describes the geodesic equation of motion \ (of a
free a particle of mass $m$) in the structure $M^{dSL}$ and the equation of
motion in that structure that must be valid if \ a particle of mass $m$ moves
in $M$ with constant bulk angular momentum $\boldsymbol{L}$. In Section 3 we
present a proof of the proposition referred above and finally in Section 4 we
present our conclusions.

\section{Preliminaries}

We now recall the description of the manifold $\mathbb{R\times}S^{3}$ as a
pseudo-sphere (a submanifold) of radius $\ell$ of the in $\mathring
{M}=\mathbb{R}^{5}$. If $(X^{0},X^{1},X^{2},X^{3},X^{4})$ are the global
Euclidian coordinates of $\mathring{M}=\mathbb{R}^{5}$ then the equation
representing the pseudo sphere is%
\begin{equation}
(X^{0})^{2}-(X^{1})^{2}-(X^{2})^{2}-(X^{3})^{2}-(X^{4})^{2}=-\ell^{2}.
\label{ds4}%
\end{equation}

Introducing \emph{conformal coordinate functions }$\{\boldsymbol{x}^{\mu}\}$
for $M$ by projecting the points of $\mathbb{R\times}S^{3}$ from the
\textquotedblleft north-pole\textquotedblright\ to a plane tangent to the
\textquotedblleft south pole" we see immediately that the
coordinates\footnote{We denote the coordinates of a point $p\in M$ covered by
the coordinate functions $\boldsymbol{x}^{\mu}$ by $x^{\mu}=\boldsymbol{x}%
^{\mu}(p)$.} $\ \{x^{\mu}\}$ covers\ all $\mathbb{R\times}S^{3}$ except the
\textquotedblleft north-pole\textquotedblright. We immediately find
that\footnote{The matrix with entries $\eta_{\mu\nu}$ is the diagonal matrix
\textrm{diag}$(1,-1,-1,-1).$}%

\begin{equation}
\boldsymbol{g}=i^{\ast}\boldsymbol{\mathring{g}}=\Omega^{2}\eta_{\mu\nu
}dx^{\mu}\otimes dx^{\nu} \label{ds1}%
\end{equation}
where%
\begin{equation}
X^{\mu}=\Omega x^{\mu},~~~~X^{4}=-\ell\Omega\left(  1+\frac{\sigma^{2}}%
{4\ell^{2}}\right)  \label{ds1a}%
\end{equation}%
\begin{equation}
\Omega=\left(  1-\frac{\sigma^{2}}{4\ell^{2}}\right)  ^{-1} \label{ds2}%
\end{equation}
and%
\begin{equation}
\sigma^{2}=\eta_{\mu\nu}x^{\mu}x^{\nu}:=x_{\mu}x^{\mu}. \label{ds3}%
\end{equation}

Now, writing $\boldsymbol{D}_{\partial_{\mu}}\partial_{\nu}=\Gamma_{\cdot
\mu\nu}^{\alpha\cdot\cdot}\partial_{\alpha}$ the non null connection
coefficients are:%
\begin{align}
{\footnotesize \Gamma}_{00}^{0}  &  {\footnotesize =}\frac{\Omega}{2l^{2}%
}{\footnotesize x}^{0}{\footnotesize ,~\Gamma}_{01}^{0}{\footnotesize =-}%
\frac{\Omega}{2l^{2}}{\footnotesize x}^{1}{\footnotesize ,~\Gamma}_{02}%
^{0}{\footnotesize =-}\frac{\Omega}{2l^{2}}{\footnotesize x}^{2}%
{\footnotesize ,~\Gamma}_{03}^{0}{\footnotesize =-}\frac{\Omega}{2l^{2}%
}{\footnotesize x}^{3}{\footnotesize ,~\Gamma}_{11}^{0}{\footnotesize =}%
\frac{\Omega}{2l^{2}}{\footnotesize x}^{0}{\footnotesize ,~\Gamma}_{22}%
^{0}{\footnotesize =}\frac{\Omega}{2l^{2}}{\footnotesize x}^{0}%
{\footnotesize ,~\Gamma}_{33}^{0}{\footnotesize =}\frac{\Omega}{2l^{2}%
}{\footnotesize x}^{0}{\footnotesize ,}\nonumber\\
{\footnotesize \Gamma}_{00}^{1}  &  {\footnotesize =-}\frac{\Omega}{2l^{2}%
}{\footnotesize x}^{1}{\footnotesize ,~\Gamma}_{01}^{1}{\footnotesize =}%
\frac{\Omega}{2l^{2}}{\footnotesize x}^{0}{\footnotesize ,~\Gamma}_{11}%
^{1}{\footnotesize =-}\frac{\Omega}{2l^{2}}{\footnotesize x}^{1}%
{\footnotesize ,~\Gamma}_{12}^{1}{\footnotesize =-}\frac{\Omega}{2l^{2}%
}{\footnotesize x}^{2}{\footnotesize ,~\Gamma}_{13}^{1}{\footnotesize =-}%
\frac{\Omega}{2l^{2}}{\footnotesize x}^{3}{\footnotesize ,~\Gamma}_{22}%
^{1}{\footnotesize =}\frac{\Omega}{2l^{2}}{\footnotesize x}^{1}%
{\footnotesize ,~\Gamma}_{33}^{1}{\footnotesize =}\frac{\Omega}{2l^{2}%
}{\footnotesize x}^{1}{\footnotesize ,}\nonumber\\
{\footnotesize \Gamma}_{00}^{2}  &  {\footnotesize =-}\frac{\Omega}{2l^{2}%
}{\footnotesize x}^{2}{\footnotesize ,~\Gamma}_{02}^{2}{\footnotesize =}%
\frac{\Omega}{2l^{2}}{\footnotesize x}^{0}{\footnotesize ,~\Gamma}_{11}%
^{2}{\footnotesize =}\frac{\Omega}{2l^{2}}{\footnotesize x}^{2}%
{\footnotesize ,~\Gamma}_{12}^{2}{\footnotesize =-}\frac{\Omega}{2l^{2}%
}{\footnotesize x}^{1}{\footnotesize ,~\Gamma}_{22}^{2}{\footnotesize =-}%
\frac{\Omega}{2l^{2}}{\footnotesize x}^{2}{\footnotesize ,~\Gamma}_{23}%
^{2}{\footnotesize =-}\frac{\Omega}{2l^{2}}{\footnotesize x}^{3}%
{\footnotesize ,~\Gamma}_{33}^{2}{\footnotesize =}\frac{\Omega}{2l^{2}%
}{\footnotesize x}^{2}{\footnotesize ,}\nonumber\\
{\footnotesize \Gamma}_{00}^{3}  &  {\footnotesize =-}\frac{\Omega}{2l^{2}%
}{\footnotesize x}^{3}{\footnotesize ,~\Gamma}_{03}^{3}{\footnotesize =}%
\frac{\Omega}{2l^{2}}{\footnotesize x}^{0}{\footnotesize ,~\Gamma}_{11}%
^{3}{\footnotesize =}\frac{\Omega}{2l^{2}}{\footnotesize x}^{3}%
{\footnotesize ,~\Gamma}_{13}^{3}{\footnotesize =-}\frac{\Omega}{2l^{2}%
}{\footnotesize x}^{1}{\footnotesize ,~\Gamma}_{22}^{3}{\footnotesize =}%
\frac{\Omega}{2l^{2}}{\footnotesize x}^{3}{\footnotesize ,~\Gamma}_{23}%
^{3}{\footnotesize =-}\frac{\Omega}{2l^{2}}{\footnotesize x}^{2}%
{\footnotesize ,~\Gamma}_{33}^{3}{\footnotesize =-}\frac{\Omega}{2l^{2}%
}{\footnotesize x}^{3}{\footnotesize .} \label{cc}%
\end{align}

Let $\sigma:I\rightarrow M,s\mapsto\sigma(s)$ be a time like geodesic in $M$.
Its tangent vector field $\sigma_{\ast}$ such that $\sigma_{\ast}%
(s)=\frac{d\boldsymbol{x}^{\mu}\circ\sigma(s)}{ds}\left.  \frac{\partial
}{\partial x^{\mu}}\right\vert _{\sigma}=\frac{dx^{\mu}}{ds}\frac{\partial
}{\partial x^{\mu}}$ satisfy $\boldsymbol{D}_{\sigma_{\ast}}\sigma_{\ast}=0$
and in components it is
\begin{equation}
\frac{d^{2}x^{\alpha}}{ds^{2}}+\Gamma_{\mu\nu}^{\alpha}\frac{dx^{\mu}}%
{ds}\frac{dx^{\nu}}{ds}=0. \label{geo1}%
\end{equation}
Using the connection coefficients given by Eq.(\ref{cc}) Eq.(\ref{geo1})
becomes
\begin{equation}
\frac{d^{2}x^{\alpha}}{ds^{2}}+\frac{\Omega}{l^{2}}x_{\mu}\frac{dx^{\mu}}%
{ds}\frac{dx^{0}}{ds}-\frac{\Omega}{2l^{2}}x^{0}\frac{dx_{\mu}}{ds}%
\frac{dx^{\mu}}{ds}=0. \label{GEOD}%
\end{equation}

\subsection{Equations of Motion in $M$ from Constant Angular Momentum in the
Bulk}

Let $\{\mathbf{E}_{A}=\frac{\partial}{\partial X^{A}}\},~A=0,1,2,3,4$ be the
canonical basis of $T\mathring{M}=T\mathbb{R}^{5}$ and let $\{E^{A}=dX^{A}\}$
be a basis of $T^{\ast}\mathring{M}$ \ dual to $\{\mathbf{E}_{A}%
=\frac{\partial}{\partial X^{A}}\}$. We have
\begin{equation}
\boldsymbol{\mathring{g}=\eta}_{AB}E^{A}\otimes E^{B} \label{MT}%
\end{equation}
where the matrix with entries $\boldsymbol{\eta}_{AB}$ is the diagonal matrix
\textrm{diag}$(1,-1,-1,-1,-1)$. Moreover let $\mathtt{\mathring{g}}=\eta
^{AB}\mathbf{E}_{A}\otimes\mathbf{E}_{B}$ be the metric of the cotangent
bundle (with $\eta^{AC}\boldsymbol{\eta}_{CB}=\delta_{B}^{A}$). Finally let
$\{E_{A}\}$ be the reciprocal basis of $\{E^{A}\},$ i.e., $\mathtt{\mathring
{g}}(E^{A},E_{B})=\delta_{B}^{A}.$ We introduce the basis $\{\mathcal{E}%
_{A}\}$ of $\mathbb{R}^{5}$ and make the usual identification $\mathbf{E}%
_{A}(p)\simeq\mathbf{E}_{A}(p^{\prime})=\mathfrak{E}_{A}$, $E_{A}(p)\simeq
E_{A}(p^{\prime})=\mathcal{E}_{A}$ for any $p,p^{\prime}\in\mathbb{R}^{5}.$

Let $\boldsymbol{X}=X^{A}\mathcal{E}_{A}$ be the position covector,
$\boldsymbol{P=}m\ddot{X}^{B}\mathcal{E}_{B}$ the bulk momentum covector and
$\boldsymbol{L}=\boldsymbol{X}\wedge\boldsymbol{P}$ the \ bulk angular
momentum of a particle of mass $m$ in the bulk spacetime $\mathbb{R}^{1,4}$.
If the particle is constrained to move "freely"\footnote{From a physical point
of view the statement moving 'freely' means that observers living in $M$
cannot detect any force acting on the particle.} in the submanifold
$\mathbb{R\times}S^{3}$ a natural hypothesis is that its bulk angular momentum
is a constant of motion. Now, $\boldsymbol{L=cte}$ implies immediately%
\begin{equation}
\frac{1}{2}(X^{A}\ddot{X}^{B}-\ddot{X}^{A}X^{B})\mathcal{E}_{A}\wedge
\mathcal{E}_{B}=0. \label{mac0}%
\end{equation}

Thus, for $\kappa,\iota=0,1,2,3$ it is $X^{\kappa}\ddot{X}^{\iota}-\ddot
{X}^{\kappa}X^{\iota}=0$, i.e.,%

\begin{equation}
x^{k}\left(  \frac{dx^{i}}{ds}\frac{1}{\ell^{2}}\Omega^{2}x_{i}\frac{dx^{l}%
}{ds}+\Omega\frac{d^{2}x^{l}}{ds^{2}}\right)  -\left(  \frac{dx^{i}}{ds}%
\frac{1}{\ell^{2}}\Omega^{2}x_{i}\frac{dx^{k}}{ds}+\Omega\frac{d^{2}x^{k}%
}{ds^{2}}\right)  x^{l}=0\label{mac}%
\end{equation}
which is the equation of motion according to the structure $M^{dSL}$.

Also, from the  equation $X^{\mu}\ddot{X}^{4}-\ddot{X}^{\mu}X^{4}=0$ we get%
\begin{gather}
-(2\Omega-1)\left(  \frac{d^{2}x^{b}}{ds^{2}}+\frac{1}{l^{2}}\Omega x_{i}%
\frac{dx^{i}}{ds}\frac{dx^{b}}{ds}\right)  \nonumber\\
+\frac{1}{2l^{2}}\Omega\left(  \frac{1}{l^{2}}\Omega x_{i}x_{j}\frac{dx^{i}%
}{ds}\frac{dx^{j}}{ds}+\frac{dx_{i}}{ds}\frac{dx^{i}}{ds}+x_{i}\frac
{d^{2}x^{i}}{ds^{2}}\right)  x^{b}=0.\label{maca}%
\end{gather}

\section{Constant Bulk Angular Momentum versus Geodesic Equation}

We have the following

\begin{proposition}
\emph{(a):} If a particle travels with geodesic motion in the structure
$M^{dSL}$ then its bulk angular momentum $\boldsymbol{L}$ is constant.
\emph{(b):} Also,if the motion of a particle of mass $m$ constrained to move
in $M$ occurs with constant bulk angular $\boldsymbol{L}$ then its motion for
an observer living in the brane $M$ is described by a timelike geodesic in the
structure $M^{dSL}$.
\end{proposition}

\textbf{Proof: }(a1) We multiply the geodesic equation for component
$x^{\kappa}(s)$ by $x^{\iota}(s)$ and the geodesic equation for component
$x^{\iota}(s)$ by $x^{\kappa}(s)$ thus getting:%
\begin{align}
\frac{d^{2}x^{\kappa}}{ds^{2}}x^{\iota}+\frac{\Omega}{\ell^{2}}x_{\mu}%
\frac{dx^{\mu}}{ds}\frac{dx^{\kappa}}{ds}x^{\iota}-\frac{\Omega}{2\ell^{2}%
}x^{\kappa}\frac{dx_{\mu}}{ds}\frac{dx^{\mu}}{ds}x^{\iota}  &  =0,\label{gk}\\
x^{\kappa}\frac{d^{2}x^{\iota}}{ds^{2}}+x^{\kappa}\frac{\Omega}{\ell^{2}%
}x_{\mu}\frac{dx^{\mu}}{ds}\frac{dx^{\iota}}{ds}-x^{\kappa}\frac{\Omega}%
{2\ell^{2}}x^{\iota}\frac{dx_{\mu}}{ds}\frac{dx^{\mu}}{ds}  &  =0. \label{gl}%
\end{align}
and subtracting Eq.(\ref{gl}) from Eq.(\ref{gk}) we get%
\begin{equation}
x^{\kappa}\left(  \frac{d^{2}x^{\iota}}{ds^{2}}+\frac{\Omega}{\ell^{2}}x_{\mu
}\frac{dx^{\mu}}{ds}\frac{dx^{\iota}}{ds}\right)  -\left(  \frac
{d^{2}x^{\kappa}}{ds^{2}}+\frac{\Omega}{\ell^{2}}x_{\mu}\frac{dx^{\mu}}%
{ds}\frac{dx^{\kappa}}{ds}\right)  x^{l}=0. \label{gkl}%
\end{equation}
Multiplying Eq.(\ref{gkl}) by $\Omega$ we get Eq.(\ref{mac}) which is the
equation of motion for the particle coming from the hypothesis that its bulk
angular momentum is constant.

(a2) From the geodesic equation (Eq.(\ref{GEOD})) we easily have the following
two equations%
\begin{align}
(2\Omega-1)\frac{d^{2}x^{k}}{ds^{2}}+(2\Omega-1)\frac{\Omega}{\ell^{2}}%
x_{i}\frac{dx^{i}}{ds}\frac{dx^{k}}{ds}-(2\Omega-1)\frac{\Omega}{2\ell^{2}%
}x^{k}\frac{dx_{i}}{ds}\frac{dx^{i}}{ds} &  =0,\nonumber\\
-\frac{1}{2\ell^{2}}\Omega x_{i}\frac{d^{2}x^{i}}{ds^{2}}x^{k}-\frac{1}%
{2\ell^{2}}\Omega\frac{\Omega}{\ell^{2}}x_{i}x_{j}\frac{dx^{j}}{ds}%
\frac{dx^{i}}{ds}x^{k}+\frac{1}{2\ell^{2}}\Omega\frac{\Omega}{2^{2}}x_{i}%
x^{i}\frac{dx_{j}}{ds}\frac{dx^{j}}{ds}x^{k} &  =0\label{mac2a}%
\end{align}
\ and thus summing these equations we get:%

\begin{gather}
(2\Omega-1)(\frac{d^{2}x^{k}}{ds^{2}})+(2\Omega-1)\frac{\Omega}{\ell^{2}}%
x_{i}\frac{dx^{i}}{ds}\frac{dx^{k}}{ds}-\frac{1}{2\ell^{4}}\Omega^{2}%
x_{i}x_{j}x^{k}\frac{dx^{i}}{ds}\frac{dx^{j}}{ds}\nonumber\\
-\frac{\Omega}{2\ell^{2}}x^{k}\frac{dx_{i}}{ds}\frac{dx^{i}}{ds}-\frac
{1}{2\ell^{2}}\Omega x_{i}(\frac{d^{2}x^{i}}{ds^{2}})x^{k}=0 \label{mac3}%
\end{gather}
\ 

Comparison of Eq.(\ref{mac3}) with Eq.(\ref{maca}). plus the result obtained
in (a1) proves that geodesic motion in the structure $M^{dSL}$ implies motion
with constant angular momentum in the bulk.\medskip

(b) Let us show now that constant bulk angular momentum $\boldsymbol{L}$
implies in geodesic motion in the structure $M^{dSL}$.

We already know that the equations of motion coming from the hypohtesis that
$\boldsymbol{L=cte}$ is for $\alpha,\beta=0,1,2,3$%

\begin{gather*}
X^{\alpha}\ddot{X}^{\beta}-\ddot{X}^{\alpha}X^{\beta}=0~~\alpha,\beta
=0,1,2,3,\\
X^{\beta}\ddot{X}^{4}-\ddot{X}^{\beta}X^{4}=0,
\end{gather*}
which can be written respectively using Eqs. (\ref{ds1a}), (\ref{ds2}) and
(\ref{ds3}) as%

\begin{equation}
x^{\alpha}(\frac{dx^{\mu}}{ds}\frac{1}{\ell^{2}}\Omega^{2}x_{\mu}%
\frac{dx^{\beta}}{ds}+\Omega\frac{d^{2}x^{\beta}}{ds^{2}})-(\frac{dx^{\mu}%
}{ds}\frac{1}{\ell^{2}}\Omega^{2}x_{\mu}\frac{dx^{\alpha}}{ds}+\Omega
\frac{d^{2}x^{\alpha}}{ds^{2}})x^{\beta}=0 \label{1}%
\end{equation}
and%

\begin{gather}
-(2\Omega-1)\left(  \frac{d^{2}x^{\beta}}{ds^{2}}+\frac{1}{\ell^{2}}\Omega
x_{\mu}\frac{dx^{\mu}}{ds}\frac{dx^{\beta}}{ds}\right) \nonumber\\
+\frac{1}{2\ell^{2}}\Omega\left(  \frac{1}{\ell^{2}}\Omega x_{\mu}x_{\nu}%
\frac{dx^{\mu}}{ds}\frac{dx^{\nu}}{ds}+\frac{dx_{\mu}}{ds}\frac{dx^{\mu}}%
{ds}+x_{\mu}\frac{d^{2}x^{\mu}}{ds^{2}}\right)  x^{\beta}=0 \label{2}%
\end{gather}

Multiplying Eq.(\ref{1}) by\ $x_{\alpha}$ (and summing in $\alpha$) we get%
\begin{equation}
x_{\alpha}x^{\alpha}(\frac{dx^{\mu}}{ds}\frac{1}{\ell^{2}}\Omega^{2}x_{\mu
}\frac{dx^{\beta}}{ds}+\Omega\frac{d^{2}x^{\beta}}{ds^{2}})-(\frac{dx^{\mu}%
}{ds}\frac{1}{\ell^{2}}\Omega^{2}x_{\mu}x_{\alpha}\frac{dx^{\alpha}}%
{ds}+\Omega x_{\alpha}\frac{d^{2}x^{\alpha}}{ds^{2}})x^{\beta}=0 \label{2a}%
\end{equation}
which can be written as%
\begin{equation}
\frac{1}{4\ell^{2}}\Omega\sigma^{2}(\frac{dx^{\mu}}{ds}\frac{1}{\ell^{2}%
}\Omega x_{\mu}\frac{dx^{\beta}}{ds}+\frac{d^{2}x^{\beta}}{ds^{2}})-\frac
{1}{4l^{2}}\Omega(\frac{1}{\ell^{2}}\Omega x_{\mu}x_{\alpha}\frac{dx^{\mu}%
}{ds}\frac{dx^{\alpha}}{ds}+x_{\alpha}\frac{d^{2}x^{\alpha}}{ds^{2}})x^{\beta
}=0 \label{3}%
\end{equation}

Summing Eq.(\ref{3}) with Eq.(\ref{2}) we get in sequence%
\[
\left(  \frac{1}{2\ell^{2}}\Omega\sigma^{2}-(2\Omega-1)\right)  \left(
\frac{1}{\ell^{2}}\Omega x_{\mu}\frac{dx^{\mu}}{ds}\frac{dx^{\beta}}{ds}%
+\frac{d^{2}x^{\beta}}{ds^{2}}\right)  +\frac{1}{2\ell^{2}}\Omega\frac
{dx_{\mu}}{ds}\frac{dx^{\mu}}{ds}x^{\beta}=0,
\]

\[
-(\frac{1}{\ell^{2}}\Omega x_{\mu}\frac{dx^{\mu}}{ds}\frac{dx^{\beta}}%
{ds}+\frac{d^{2}x^{\beta}}{ds^{2}})+\frac{1}{2\ell^{2}}\Omega\frac{dx_{\mu}%
}{ds}\frac{dx^{\mu}}{ds}x^{\beta}=0,
\]

and finally$\frac{d^{2}x^{\beta}}{ds^{2}}+\frac{1}{\ell^{2}}\Omega x_{\mu
}\frac{dx^{\mu}}{ds}\frac{dx^{\beta}}{ds}-\frac{1}{2\ell^{2}}\Omega x^{\beta
}\frac{dx_{\mu}}{ds}\frac{dx^{\mu}}{ds}.$which is just the geodesic equation
(see Eq.(\ref{GEOD})) in the structure $M^{dSL}$.$\blacksquare$

\section{Conclusions}

We said in the introduction that the de Sitter structure $M^{dSL}$ has been
studied by many authors as a possible natural arena for the motion of
particles and fields instead of the Minkowski spacetime structure
$\mathfrak{M}$. In particular papers \cite{ps,pss} are devoted to study which
could be the natural motion of particles in $M^{dSL}$. Authors of that papers
claim that the natural path of free particles in $M^{dSL}$ cannot be timelike
geodesics in that structure and they build a new theory for finding the true
geodesics for the structure $(M,\boldsymbol{g}).$ Their main argument of such
a claim is that they became confused by a quotation in \cite{he} saying that
there are points in $M$ which cannot be joined by a geodesic. However, the
quotation in \cite{he} is only partially valid. Indeed, it has been long
showed in \cite{schmidt} that points that cannot be joined by a geodesic can
only be joined by a spacelike curve, which of course cannot be the paths of
any material particle. We discussed these issues at length in \cite{rw2015}
where in particular it is shown also that he equations of motion found in
\cite{ps,pss} are equivocated. At least we want to emphasize that recently it
has been shown in \cite{r+} by using the Clifford and spin-Clifford formalisms
\cite{rodcap2007} that the hypothesis that a particle moving freely in
$(M,\boldsymbol{g)}$ has constant bulk angular momentum leads naturally to the
Dirac equation as found in \cite{dirac} in the de Sitter structure
$(M,\boldsymbol{g})$.

\end{document}